\documentclass[a4paper,12pt]{article}
\usepackage[polish, english]{babel}
\usepackage[cp1250]{inputenc}
\usepackage{amsmath,amsmath,graphicx,epsf,pstricks,amsfonts,epsfig,multicol}
\usepackage[T1]{polski}
\usepackage[mathscr]{eucal}
\usepackage{titlesec}
\titleformat{\section}
  {\normalfont\Large\bfseries\centering}{\thesection}{1em}{}
\titleformat{\subsection}
  {\normalfont\large\bfseries\centering}{\thesubsection}{1em}{}

\numberwithin{equation}{section}

\title{\textbf{Induced gauge interactions revisited}}
\author{Bogus{\l}aw
Broda\footnote{bobroda@uni.lodz.pl}\; and Micha{\l} Szanecki\footnote{michalszanecki@wp.pl}\\
\small\textit{Department of Theoretical Physics,}
\small\textit{University of {\L}\'od\'z}\\
\small\textit{Pomorska 149/153,\;90-236 {\L}\'od\'z, Poland}\\
}

\begin{document}
\maketitle
\begin{abstract}
\noindent
\begin{quote}
It has been shown that the old-fashioned idea of Sakharov's
induced gravity and gauge interactions in the ``one-loop
dominance" version works astonishingly well yielding reasonable
parameters. It appears that induced coupling constants of gauge
interactions of the standard model assume qualitatively realistic
values. Moreover, it is possible to induce the Barbero--Immirzi parameter of canonical gravity from the fields entering the standard model.\\\\

\textbf{PACS number(s):} 11.15.Tk Other nonperturbative
techniques; 04.62.+v Quantum fields in curved spacetime; 12.10.Dm
Unified theories and models of strong and electroweak
interactions; 02.40.Hw Classical differential geometry; 02.40.Ma
Global differential geometry; 04.20.Fy Canonical formalism,
Lagrangians, and variational principles; 04.60.Pp Loop quantum
gravity, quantum geometry, spin foams; 04.70.Dy Quantum aspects of
black holes, evaporation, thermodynamics; 11.30.Rd Chiral
symmetries; 12.60.Rc Composite models.
% KEYWORDS:
% induced gravity; induced interactions; black-hole entropy; heat-kernel method; Schwinger proper-time method.
\end{quote}
\end{abstract}
%\begin{multicols}{2}
\eject
\section{Introduction}
\noindent The idea that fundamental interactions might be not so
fundamental as they appear, but induced by quantum fluctuations of
the vacuum emerges from the fifties of the 20th century when in
1967 Sakharov and Zel'dovich published their papers about induced
gravity \cite{Sakharov} and induced electrodynamics
\cite{Zeldovich}. After a while it was followed by many people
(see, e.g.\ \cite{Birula}) and firstly applied to the context of
the standard model by Terazawa \cite{Terazawa1}, who also derived
logarithmic relation between the gauge coupling constants and the
Newton gravitational constant \cite{Terazawa2} (see also
\cite{Landau}). Some further, explorations have been presented in
\cite{Akama3},\cite{Akama4}.

It seems that successful application of the idea of quantum vacuum
induced interactions to the two fundamental interactions
subsequently renewed interest in this subject. At present, the
very idea lacks a clear theoretical interpretation. It is supposed
to be an interesting curiosity as well as an unexplained deeper
phenomenon. Despite of an interpretation, striking coincidences
have forced us to claim that the idea of quantum induced
interactions does work and can be used practically.

The aim of our paper is to show that the idea of induced gauge
interactions in its primary, old-fashioned, ``one-loop dominance''
Sakharov's version  \cite{Visser} yields phenomenologically very
realistic results. This standpoint assumes that in the classical
action at the beginning there are only (fundamental) matter fields
coupled (minimally) to external gauge fields. Classical terms for
gauge fields and gravity do not exist autonomously, but they
appear by virtue of the low-order one-loop calculations. (The
superior role of the matter fields awaits an explanation in this
framework.) Interestingly, classical terms produced this way have
not only appropriate functional term but also realistic numeric
coefficients. The former fact is akin to renormalizability.

In our paper, gauge interactions and gravity in the modern
Ashtekar's canonical formulation are treated and analyzed in the
framework of the very convenient method: Schwinger's proper time
and the Seeley--DeWitt heat-kernel expansion.
Our approach to the gauge interactions uses flat Lorentzian
manifold whereas for the Ashtekar's gravity our calculations are
carried out on the Riemannian one with torsion (extension
including torsion has been given in \cite{DS3}), so that the
one-loop effective actions for each case are different and need
individual treatment.

\eject

\section{Induced gauge interactions}
\subsection{Heat-kernel method}
\noindent According to the idea of quantum vacuum induced
interactions, its dynamics emerges from dominant contributions to
the one-loop effective action of non-self-interacting matter
fields coupled to these interactions. In the framework of the
Schwinger proper-time method, the expected terms for gauge
interactions can be extracted from the 2nd coefficient of the
Seeley--DeWitt heat-kernel expansion. (``Cosmological constant''
and gravity are obtainable from the 0th and 1st coefficient
respectively, see our work \cite{Broda1}.) In Minkowskian
signature \cite{Birrell},\cite{DeWitt}
\begin{align}
 S_{\rm eff}=i\kappa\log\det\mathcal{D}=i\kappa\mathrm{Tr}\log\mathcal{D}=-i\kappa\int\frac{\mathrm{d}s}{s}\;
 \mathrm{Tr}\;e^{-is\mathcal{D}},
 \label{eq:EffectiveAction1}
\end{align}
where $\mathcal{D}$ is an appropriate second-order differential
operator, and $\kappa$ depends on the kind of the ``matter'' field
(its statistics, in principle). E.g.\ for a bosonic single mode,
$\kappa=\frac{1}{2}$. Seeley--DeWitt heat-kernel expansion (for
detailed introduction, see Appendix) in four dimensions, reads
\begin{align}
\mathrm{Tr}\;e^{-is\mathcal{D}}=\frac{1}{16{\pi}^{2}\left(is\right)^2}\left[A_{0}+A_{1}\left(is\right)+A_{2}\left(is\right)^2+\cdots\right],
\label{eq:Seeley-DeWiitExpansion1}
\end{align}
where $A_{n}$ is the $n$th Seeley--DeWitt coefficient. Imposing
appropriate cutoffs, i.e.\ an UV cutoff $\varepsilon$ for $A_0$,
$A_1$ and $A_2$, and an IR cutoff $\Lambda$ for $A_2$, we obtain
\begin{align}
S_{\rm
eff}=\frac{\kappa}{16{\pi}^{2}}\left(\frac{1}{2}A_{0}\varepsilon^{-2}+A_{1}\varepsilon^{-1}+A_{2}\log\frac{\Lambda}{\varepsilon}+\cdots\right).
\label{eq:Seeley-DeWiitExpansion2}
\end{align}
As we stated earlier, we are especially interested in the gauge
part connected with the $A_2$. Collecting contributions from
various modes, we get the following gauge Lagrangian density:
\begin{align}
\mathcal{L}_{2}=\frac{1}{384{\pi}^{2}}\log\frac{\Lambda}{\varepsilon}\left(N_{0}+4N_{\frac{1}{2}}\right)\mathrm{tr}F^{2}
\label{eq:LagrangeDensity2}
\end{align}
(see, Table 2 in Appendix for the origin of the numeric
coefficient), where:
\begin{equation}
\begin{split}
&N_0=\mbox{number of minimal scalar degrees of freedom (dof),}\\
&N_{\frac{1}{2}}=\mbox{number of two-component fermion fields}=\mbox{half the number of fermion dof,}\\
\end{split}
\label{Notations}
\end{equation}
and $F$ is the strength of a gauge field (see, the definition
\eqref{eq:A.1}.) Higher-order terms are in principle present (even
in classical case), but they are harmless in typical situations
because of small values of the coefficients following from the
cutoffs.

\subsection{Standard model contributions}
Now, we will concentrate on the possibility of quantum generation
of gauge interactions in the context of the standard model. We are
interested in the contributions coming from appropriate matter
fields taken into account in \eqref{eq:LagrangeDensity2}. To this
end we should adapt \eqref{eq:LagrangeDensity2} to the context of
the standard model. Adopting the matter contents of the Lagrangian
of the standard model we display all contributions to the
respective gauge parts in Table 1.
%\begin{center}
\begin{figure}\label{Table 1}
\includegraphics[scale=0.95]{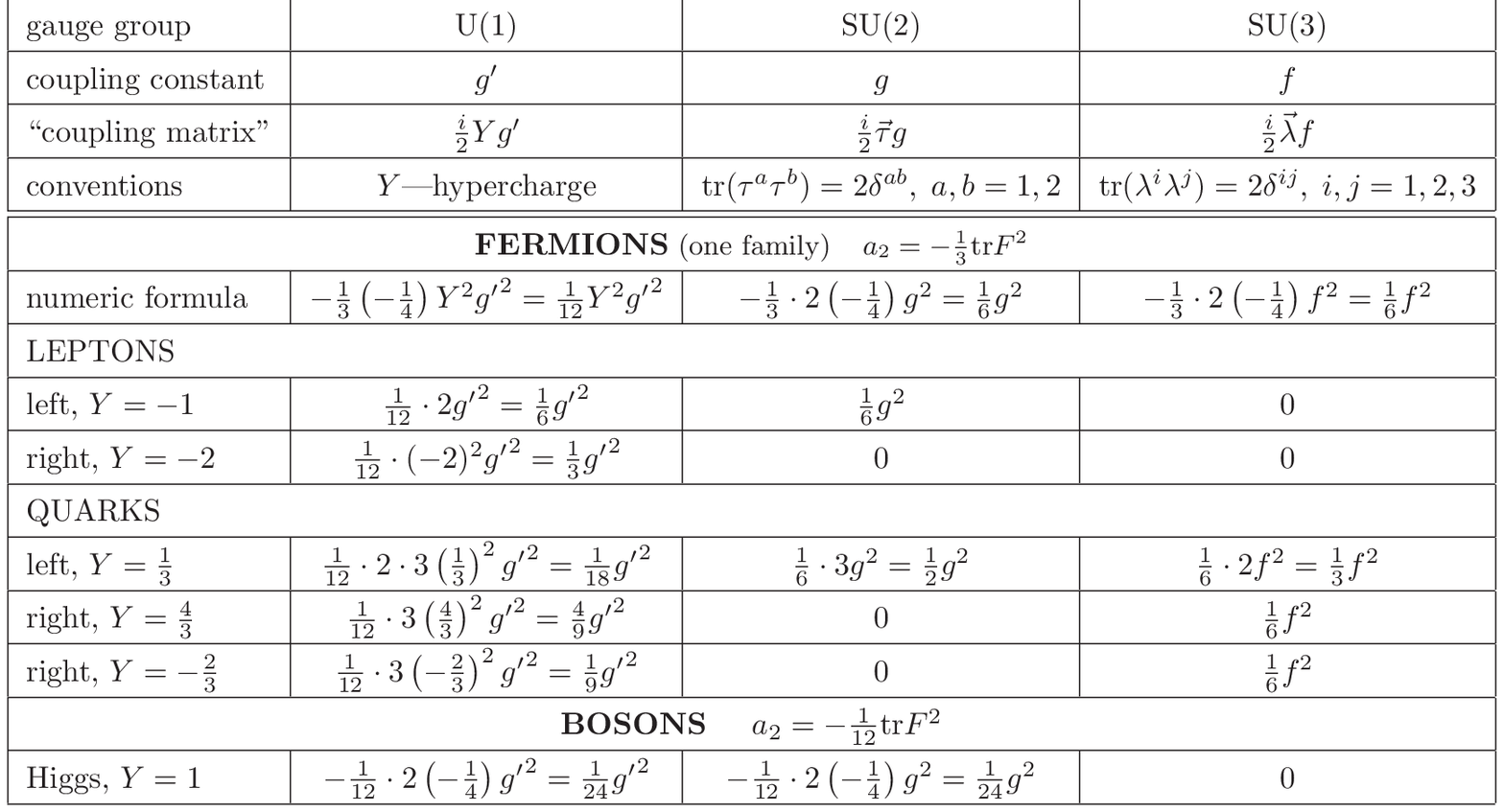}
\begin{quote}
\textit{\textbf{Table 1}: Various ``matter field'' species
contributions to induced gauge coupling constants in the framework
of the standard model.}
\end{quote}
\end{figure}
The assumed implicit convention for the operator of covariant
derivative is
\begin{align}
D_{\mu}=\partial_{\mu}+\vec{X}\cdot\vec{A}_{\mu},
\label{eq:Covariant1}
\end{align}
where $\vec{X}$ is the ``coupling matrix'' given in the third row
of Table 1. Strictly speaking, $\vec{X}$ is a tensor product with
two matrix units corresponding to the other two gauge groups,
yielding additional coefficients, 2 or 3. In principle, the
coefficients given in each column and multiplied by
\begin{align}
\frac{1}{16{\pi}^2}\log\frac{M}{m}, \label{Coefficient1}
\end{align}
where $M$ and $m$ is an UV and an IR cutoff, respectively, in mass
units, should sum up to $\frac{1}{4}$, a standard normalization
term in front of $F^2$, so that, we have the following theoretical
constraint:
\begin{align}
\frac{{g_{i}}^2}{16{\pi}^2}\sum_{n}\alpha_{(i)n}\log\frac{M_n}{m_n}=\frac{1}{4},
\label{Coefficient2}
\end{align}
where $g_i\;(i=1,2,3)$ is one of the three coupling constants,
$\alpha_{(i)n}$ are corresponding numeric coefficients from Table
1, and the sum concerns all matter fields. We can confidently set
$M_n=M_{\rm P}$ (Planck mass) and $m_n$ are masses of lightest
particles.

Now, we can utilize the data given in Table 1 in order to
reproduce a number of phenomenologically realistic results.
Assuming for simplicity (or as an approximation) fixed values of
$M_n$ and $m_n$ for all species of matter particles, we can
uniquely rederive following \cite{Terazawa} the Weinberg angle
$\theta_{\rm \textsc{w}}$,
\begin{align}
\sin^2\theta_{\rm
\textsc{w}}=\frac{{g'}^2}{g^2+{g'}^2}\approx0.38\;.
\label{eq:WeinbergAngle}
\end{align}
Estimation of the coupling constants requires definite values of
infrared cutoffs $m_n$. For $m_n$ of the order of the mass of
lighter particles of the standard model we obtain
\begin{align}
\alpha=\frac{e^2}{4\pi}=\frac{g^2\sin^2\theta_{\rm
\textsc{w}}}{4\pi}=\mathcal{O}(0.01), \label{eq:SubtleConstant}
\end{align}
and
\begin{align}
g=f=\mathcal{O}(1),
 \label{eq:Coupling}
\end{align}
which is phenomenologically a very realistic estimate.

Alternatively, the constraint \eqref{Coefficient2} can give some
limitations on the ratio of the two scales $M$ and $m$, provided
the scale of interactions $g=f=\mathcal{O}(1)$ is assumed.

\section{Ashtekar gravity: the Barbero--Immirzi parameter}
\subsection*{Introduction}
The Barbero--Immirzi (BI) parameter $\gamma$ is an a priori free
parameter in the framework of the modern approach to canonical
gravity (Ashtekar's formalism) \cite{Ashtekar}. In the Holst
extended action for gravity \cite{Holst} the BI parameter $\gamma$
resides in the additional term of the full (Holst) action. One can
easily further extend the Holst contribution \cite{Rezende}
yielding, in particular, the Nieh--Yan (NY) term \cite{Nieh}. (The
role that NY invariant plays in gravity has been studied in
\cite{Mercuri1}, while an extension to a possible new scenario
where BI parameter is promoted to a field, has been studied in
\cite{Mercuri2}). Topological nature of the NY term, means that it
influences quantum theory. Interestingly, it appears, and we will
show it, that, the  NY term can be quantumly induced by dominant
part of one-loop contributions coming from chiral matter fields.

\subsection{Formalism}
We will work in the framework of euclidean formalism applying the
Sakharov idea of induced gravity (one-loop dominance) to the
standard model of particle physics. We are especially interested
in showing that chiral fields entering the standard model
(left-handed leptons, i.e.\ neutrinos, in our case) will yield an
additional term \cite{BroSza1}, the NY term. Consequently, the
induced BI parameter depends only on the number and kind of
particle species entering the standard model. From purely
technical point of view the calculus is partially akin to the
derivation of the Adler--Bell--Jackiw (ABJ) chiral anomaly in
space-time with torsion \cite{Zanelli} (see \cite{Mielke1} for an
extended discussion on the anomaly issue).

According to our realization of the Sakharov idea, we are
interested in a dominant part of one-loop contributions coming
from left-handed leptons. We will work in the (euclidean)
Schwinger proper-time formalism and in the framework of the
Seeley--DeWitt heat-kernel expansion on manifolds with torsion
\cite{Obukhov}. Our starting object is the Dirac differential
operator \begin{align} D\equiv i\! \not\!\nabla\equiv
i\gamma^{a}e^{\mu}_{a}\,\nabla_{\mu}, \label{eq: Dirac
differential operator} \end{align} where $e^{\mu}_{a}$ is a
vierbein field, $\nabla_{\mu}$ is a covariant derivative in space
with torsion, and $\gamma^{a}$ are euclidean Dirac matrices. Now
\begin{align}
D^{2}=-\square+\frac{1}{2}e^{\mu}_{a}e^{\nu}_{b}\sigma^{ab}T^{\lambda}_{\mu\nu}\nabla_{\lambda}-\frac{1}{8}e^{\mu}_{a}e^{\nu}_{b}\sigma^{ab}\sigma^{cd}R_{cd\mu\nu},
\label{eq: Dirac differential operator squared} \end{align} where
\begin{align}
\square\equiv\nabla_{\mu}\nabla^{\mu},\;\;\sigma^{ab}\equiv\frac{1}{2}\left[\gamma^{a},\gamma^{b}\right],\;\;\left[\nabla_{\mu}\,
,\nabla_{\nu}\right]V^{a}=R^{a}_{\;b\mu\nu}V^{b}-
T^{\lambda}_{\mu\nu}\nabla_{\lambda}V^{a}. \label{eq: Dirac
differential operator squared-notation} \end{align} Introducing
the two chiral projectors \begin{align} P_{\rm
\textsc{l}}\equiv\frac{1-\gamma^{5}}{2},\;\;P_{\rm
\textsc{r}}\equiv\frac{1+\gamma^{5}}{2}, \label{eq: Projectors
with gamma5} \end{align} with
$\gamma^{5}\equiv\gamma^{1}\gamma^{2}\gamma^{3}\gamma^{4}$, we can
write in the chiral representation \begin{align} D^{2}=D^{2}P_{\rm
\textsc{l}}\oplus D^{2}P_{\rm \textsc{r}}, \label{eq: Dirac
differential operator squared-simple sum}
\end{align} and consequently \begin{align} \det D=\sqrt{\det
D^{2}}=\sqrt{{\det}_{\rm \textsc{l}} D^{2}\,{\det} _{\rm
\textsc{r}} D^{2}}, \label{eq: Dirac differential operator
determinant} \end{align} because $D^{2}$ is diagonal-blocked in
the subspaces L and R. From now on we will confine ourselves to
$\sqrt{{\det}_{\rm \textsc{l}} D^{2}}$ corresponding to the
left-handed lepton.\\

\subsection{Effective action}
The effective action for the chiral (left-handed) lepton is of the
following form
\begin{align}
S=-\frac{1}{2}\log{\det}_{\rm\textsc{l}}
D^{2}=\frac{1}{2}\int\frac{\mathrm{d}s}{s}\mathrm{Tr}\left(e^{-sD^{2}}P_{\rm
\textsc{l}}\right). \label{eq: Effective action for the chiral
lepton}
\end{align}
Then, corresponding chiral part of the $M^2$-regularized effective
Lagrangian density reads (see, \cite{Zanelli})
\begin{align}
\mathcal{L}=\frac{1}{2}\int\limits_{M^{-2}}^{\infty}\frac{\mathrm{d}s}{s}\,\frac{s}{(4\pi
s)^{2}}\,\left(-\frac{1}{2}\right)\mathrm{tr}\left(a_{1}\gamma^{5}\right)\approx-\frac{1}{4}\left(\frac{M}{4
\pi}\right)^{2}\mathcal{NY}, \label{eq: Effective Lagrangian for
the chiral lepton}
\end{align}
where $a_{1}$ is the 1st Seeley--DeWitt coefficient
\cite{Obukhov}, and the NY term $\mathcal{NY}$ is defined by
\begin{align}
\mathcal{NY}\equiv\mathrm{d}_{\omega}e^{a}\wedge\mathrm{d}_{\omega}e_{a}-e^{a}\wedge
e^{b}\wedge R_{ab}\equiv T^{a}\wedge T_{a}-e^{a}\wedge e^{b}\wedge
R_{ab}. \label{eq: Nieh-Yan term definition}
\end{align}
The extended Lagrangian density of general relativity assumes the
form
\begin{align} \mathcal{L}=\alpha\star\left(e^{a}\wedge
e^{b}\right)\wedge R_{ab}-\beta\left(T^{a}\wedge T_{b}-e^{a}\wedge
e^{b}\wedge R_{ab}\right), \label{eq: Extended Lagrangian}
\end{align}
where the first term is the standard Einstein--Hilbert (EH) one,
and the second term is the extended Holst or the NY one. The
Barbero--Immirzi parameter $\gamma$ is now given by
\begin{align}
\gamma\equiv\frac{\alpha}{\beta}. \label{eq: Barbero-Immirzi
parameter}
\end{align}
Using the result of \cite{Broda1} we have
\begin{align} \mathcal{L}_{\rm
\textsc{eh}}=-\frac{1}{12}\,\left(\frac{M}{4\pi}\right)^{2}\left(N_{0}+N_{\frac{1}{2}}-4N_{1}\right)\star\left(e^{a}\wedge
e^{b}\right)\wedge R_{ab}, \label{eq: E-H Lagrangian} \end{align}
where $N_{0}$, $N_{\frac{1}{2}}$ are defined by \eqref{Notations},
and $N_{1}$ is the number of gauge fields. Therefore, by virtue of
\eqref{eq: Effective Lagrangian for the chiral lepton}, \eqref{eq:
Nieh-Yan term definition} and \eqref{eq: Extended
Lagrangian}--\eqref{eq: E-H Lagrangian}
\begin{align}
\gamma=\frac{-\frac{1}{12}\left(N_{0}+N_{\frac{1}{2}}-4N_{1}\right)}{-\frac{1}{4}N_{\rm
\textsc{l}}}, \label{eq: Induced gamma parameter} \end{align}
where $N_{\rm \textsc{l}}$ is the number of chiral left-handed
modes, and the UV cutoffs $\left(M/4 \pi \right)^{2}$ canceled
out in \eqref{eq: Induced gamma parameter}.\\ For example, exactly
in the framework of the standard model, we insert the following
numbers of fundamental modes: $N_{0}=4$ (Higgs),
$N_{\frac{1}{2}}=45$, $N_{1}=12$, $N_{\rm \textsc{l}}=3$
(neutrinos), yielding $\gamma=\frac{1}{9}\approx 0.11$, which is
quite close to the (a bit obsolete)
Ashtekar--Baez--Corichi--Krasnov value,
$\gamma_{\!\rm\textsc{abck}}=\frac{\ln2}{\pi\sqrt{3}}\approx 0.13$
\cite{Baez},\cite{Domagala} (see \cite{Meissner}, for a better
estimation). Nevertheless we should remember that $\gamma$ induced
that way depends on the number and kinds of fundamental modes, and
moreover the whole calculus is valid in the framework of euclidean
formalism.\\
One should stress that the result
$\gamma_{\!\rm\textsc{abck}}\approx 0.13$ is obtained in the
framework of an approach using the black-hole entropy. We have
shown, and this is the main objective of our considerations, that
our method of the Sakharov's induced NY term also fix the BI
parameter $\gamma$, and moreover it does it in an independent way.

\section{Final remarks}
\noindent In this paper, we have presented a number of arguments
supporting the idea of the old-fashioned ``one-loop dominance''
version of gauge interactions in the spirit of Sakharov. All
coupling constants of fundamental gauge interactions have been
shown to assume phenomenologically realistic values, provided the
Planckian value of the UV cutoff is given. As an independent check
of the idea, we have proposed Sakharov's approach to the modern
canonical Ashtekar's gravity. The a priori free Barbero--Immirzi
parameter, by virtue of the chiral fields contributions coming
from the standard model, also assume an acceptable value in the
euclidean framework.

\section*{Acknowledgments}
This work was supported in part by the University of {\L}\'od\'z
grant.

\eject
\begin{appendix}
\section*{Appendix}
\section{Heat-kernel expansion} The functional trace in
Eq.~\eqref{eq:Seeley-DeWiitExpansion1}, by definition reads:
\begin{align}
\mathrm{Tr}\;e^{-is\mathcal{D}}\equiv\int\limits_{\mathcal{M}}\mathrm{d}x\,\mathrm{tr}\left<x\right|e^{-is\mathcal{D}}\left|x\right>,
\end{align}
where ``$\mathrm{tr}$'' is an ordinary algebraic trace. In order
to calculate this integrand, let us consider the, so called,
heat-kernel, defined by:
\begin{align}
\mathscr{G}(x,y;s)\equiv\left<x\right|e^{-s\mathcal{D}}\left|y\right>.
\label{eq: jadro przewodnictwa 1}
\end{align}
It is easy to check, that non-interacting version of the
heat-kernel equation assumes the form:
\begin{align}
\frac{\partial \mathscr{G}_{0}(x,y;s)}{\partial
s}=\partial^{2}\mathscr{G}_{0}(x,y;s), \label{Rownanie
przewodnictwa dla jadra-teoria swobodna}
\end{align}
where
\begin{align}
\mathscr{G}_{0}(x,y;s)=\left<x\right|e^{-s\mathcal{D}_{0}}\left|y\right>
\end{align}
is defined by a free operator $\mathcal{D}_{0}$. Consequently,
solving out Eq.~\eqref{Rownanie przewodnictwa dla jadra-teoria
swobodna}, in $d$ dimensions, we have:
\begin{align}
\mathscr{G}_{0}(x,y;s)=\frac{1}{(4 \pi
s)^{d/2}}e^{-\left|x-y\right|^{2}/4s}. \label{Rownanie
przewodnictwa dla jadra-teoria swobodna 1}
\end{align}
Making use of a perturbation techniques with respect to $s$, we
can write the final solution for a general operator $\mathcal{D}$:
\begin{align}
\mathscr{G}(x,y;s)=\frac{1}{(4 \pi
s)^{d/2}}e^{-\left|x-y\right|^{2}/4s}F(x,y;s), \label{Rownanie
przewodnictwa dla jadra-teoria ogolna-rozwiazanie-cd 3}
\end{align}
where  $F(x,y;s)$ is a matrix valued function represented by the
perturbative expansion:
\begin{align}
F(x,y;s)\equiv\sum\limits_{j=0}^{\infty}A_{j}(x,y)s^{j}.
\label{Rozwiniecie potegowe funkcji F-heat kernel expansion 1}
\end{align}
The $A_{j}(x,y)$ coefficients are Seeley--DeWitt (Hamidew)
coefficients widely explored in numerous papers, e.g.\
\cite{Ball},\cite{Vassilevich}. Recapitulating, going back to
Eq.~\eqref{eq: jadro przewodnictwa 1} with $x=y$ condition, for
$d=4$ in the relativistic version (Wick's rotation), we finally
obtain Eq.~\eqref{eq:Seeley-DeWiitExpansion1}.

\section{Seeley--DeWitt coefficients}
The Seeley--DeWitt (``Hamidew'') coefficients assume the values
presented below (in Table 2). Our sign convention corresponds to
the Landau--Lifshitz timelike one, i.e.\ the metric signature is
$\left(+---\right)$. Our conventions concerning gauge fields are
as follows:
\begin{equation}
\begin{split}
&D_{\mu}=\partial_{\mu}+A_{\mu},\\
&F_{\mu\nu}=\partial_{\mu}A_{\nu}-\partial_{\nu}A_{\mu}+\left[A_{\mu},A_{\nu}\right].
\end{split}
\label{eq:A.1}
\end{equation}
\\

\begin{center}
 \begin{tabular}{|c||c||}
  \hline
  particle & {Seeley--DeWitt coefficient $k_2$}\\
  \cline{1-2}
  \hline \hline
  minimal scalar & $\frac{1}{12}$\; \cite{DeWitt}\\
  \hline
  Weyl spinor & $-\frac{1}{3}$\;\cite{DeWitt}\\
  \hline
  massless vector & $-\frac{11}{24}$\; \cite{Vassilevich}\\
  \hline
 \end{tabular}
\end{center}
\begin{quote}
\textit{\textbf{Table 2}: Seeley--DeWitt coefficients. In
brackets, we have given the references where the coefficients can
be found explicitly or almost explicitly (i.e.\ after few-minute
calculations).}
\end{quote}
We have assumed the following notation:
\begin{equation}
a_{2}(x)=k_{2}\,{\rm{tr}}F^{2}+k'_{2} \textrm{``curvature
terms''}. \label{eq:A.2}
\end{equation}
Interested reader can find $k'_2$ in \cite{Visser}, \cite{Birrell}
or \cite{Vassilevich}.
\end{appendix}

%\begin{center}
%\includegraphics[scale=0.82]{tabela}
%\end{center}

%\end{multicols}
\end{document}